\documentclass[11pt,a4paper]{article}
\usepackage{amsmath,amssymb}
\usepackage{graphicx}

\newcommand{\lan}{\langle}
\newcommand{\ran}{\rangle}

\newcommand{\rr}{\mathbf{r}}
\newcommand{\pp}{\mathbf{p}}

\newcommand{\xxi}{\mathcal{F}}

\begin{document}

\title{\bf The coupled-channel analysis of $D_s$ and $B_s$ mesons}
\author{A.~M.~Badalian\footnote{{\bf e-mail}: badalian@itep.ru},
Yu.~A.~Simonov\footnote{{\bf e-mail}: simonov@itep.ru},
M.~A.~Trusov\footnote{{\bf e-mail}: trusov@itep.ru}
\\
 \small{\em Institute of theoretical and experimental physics} \\
\small{\em Russia 117218, Moscow, Bolshaya Cheremushkinskaya str.,
25} }
\date{}
\maketitle

\begin{abstract}
In the framework of the coupled channel model the mass shifts of
the $P$--wave excitations of $D_s$ and $B_s$ mesons have been
calculated. The corresponding coupling to $DK$ and $BK$ channels
is provided by the effective chiral Lagrangian which is deduced
from QCD and does not contain fitting parameters. The strong mass
shifts down for $0^+$ and ${1^+}'$ states have been obtained,
while ${1^+}''$ and $2^+$ states remain almost at rest. Two
factors are essential for large mass shifts: strong coupling of
the $0^+$ and $1^{+'}$ states to the $S$-wave decay channel,
containing a Nambu-Goldstone meson, and the chiral flip
transitions due to  the bispinor structure of both heavy-light
mesons. The masses $M(B^*_s(0^+))=5710(15)$ MeV and
$M(B_s(1^{+'}))=5730(15)$ MeV are predicted. Experimental limit on
the width $\Gamma(D_{s1}(2536))<2.3$ MeV puts strong restrictions
on admittable mixing angle between the $1^+$ and $1^{+'}$ states.
\end{abstract}

\section{Introduction}

The heavy-light (HL) mesons  play a special role in hadron
spectroscopy. First of all, a HL meson is the simplest system,
containing one light quark in the field of almost static heavy
antiquark, and that allows to study quark (meson) chiral
properties. The discovery of the $D_s(2317)$ and $D_s(2460)$
mesons \cite{1,2} with surprisingly small widths and low  masses
has given an important impetus to study chiral dynamics and raised
the question why their masses are considerably lower than expected
values in single channel potential models. The question was
studied in different approaches: in relativistic quark model
calculations \cite{3I}--\cite{6I},  on the lattice \cite{7I}, in
QCD Sum Rules \cite{8I,9I}, in chiral models
\cite{10I}--\cite{12I} (for reviews see also \cite{13I,14I}). The
masses of $D_s(0^+)$ and $D_s(1^{+'})$ in closed-channel
approximation typically exceed by $\sim$ 140 and 90 MeV their
experimental numbers.

Thus  main theoretical goal  is  to understand dynamical mechanism
responsible for such large mass shifts of the $0^+$ and $1^{+'}$
levels (both states have  the light quark orbital angular momentum
$l=0$ and $j=1/2$) and   explain why the position of
 other two levels (with $j=3/2)$ remains practically unchanged.
The importance of second fact has been underlined by S.Godfrey in
\cite{5I}.

 The mass shifts of the $D_s(0^+,1^{+'})$ mesons have already been
considered in a number of papers with the use of unitarized
coupled-channel model \cite{15I}, in nonrelativistic Cornell model
\cite{16I}, and in different chiral models \cite{17I}--\cite{19I}.
Here we address again this problem with the aim to calculate also
the mass  shifts of the $D_s(1^{+''})$ and $B_s(0^+,1^{+'})$
states and the widths of the $2^+$ and $1^+$ states,  following
the approach developed in \cite{18I}, for which strong coupling to
the S-wave decay channel, containing a pseudoscalar ($P$)
Nambu-Goldstone (NG) meson, is crucially important. Therefore in
this approach principal difference exists between vector-vector
($VV$) and $VP$ (or $PP$) channels. This analysis of two-channel
system is performed with the use of the chiral quark-pion
Lagrangian which has been  derived directly from the QCD
Lagrangian \cite{20I} and   does not contain fitting parameters,
so that the shift of the $D^*_s(0^+)$ state $\sim$ 140 MeV is only
determined by the conventional decay constant $f_K$.

Here the term ``chiral dynamics'' implies the mechanism by which
in the transition from one HL meson to another the octet of the NG
mesons $\phi$ is emitted. The corresponding Lagrangian $\Delta
L_{FCM}$,
\begin{equation}\Delta L_{FCM}=\bar q (\sigma r ) \exp (i\gamma_5
\phi/f_\pi)q,\label{1.1}\end{equation}contains the light-quark
part, $\exp (i\gamma_5 \phi/f_\pi)$, where $\phi$ is the $SU(3)$
octet of NG mesons and the important factor $\gamma_5$ is present.
In the lowest order in $\phi$ this Lagrangian coincides with
well-known effective Lagrangian $\Delta L_{\text{eff}}$ suggested
in \cite{21I},\cite{22I}, where, however, an arbitrary constant
$g_A$ is introduced . At large $N_c$, as argued in \cite{21I},
this constant has to be equal unity, $g_A=1$. In
\cite{10I,17I,22I} this effective Lagrangian was applied to
describe decays of HL mesons taking $g_A<0.80$.

More general Lagrangian $\Delta L_{FCM}$ (\ref {1.1}) was derived
in the framework of the field  correlator method (FCM) \cite{20I,
23I}, in which the constant $g_A=1$ in all cases, and which
contains NG mesons to all orders, as seen from its explicit
expression (\ref{1.1}).

In \cite{BST} with the use of the Dirac equation it was shown that
in the lowest order in $\phi$   $\Delta L_{FCM}=\Delta
L_{\text{eff}}$, if  indeed $g_A=1$. In our calculations the
$\Delta L_{FCM}$ was used to derive the nonlinear equation for the
energy shift  and width, $\Delta E=\Delta\bar
E-\frac{i\Gamma}{2}$, as in \cite{18I}. We do not assume any
chiral dynamics for the unperturbed levels, which are calculated
here with the use of the QCD string Hamiltonian \cite{24I,25I},
because the mass  shift $\Delta E$ appears to be weakly dependent
on the position of unperturbed level. Nevertheless, the
uncertainty in the final mass values is due to a poor knowledge of
the fine structure (FS) interaction in the initial (unperturbed)
P-wave masses.

It is essential that resulting shifts of the $J^P(0^+,1^{+'})$
levels are large only for the $D_s,B_s$ mesons, which lie close to
the $DK,D^*K,BK,B^*K$ thresholds, but not for the $D(1P),B(1P)$
mesons, in this way violating symmetry between them (this symmetry
is possible in close-channel approximation). In our calculations
 shifted  masses of the $D_s(0^+)$  and
$B_s(0^+)$ practically coincide  with those for the $D(0^+)$ and
$B(0^+)$, in agreement with the experimental fact that
$M_{\exp}(D(0^+))= 2350\pm 50$ MeV \cite{26I} is equal or even
larger than $M_{\exp}(D_s(0^+))=2317$ MeV. The states with
$j=3/2~~ D_s(1^+,2^+)$ and $D(1^+,2^+)$ have no mass shifts and
for them the mass difference is $\sim 100$ MeV, that just
corresponds to the mass difference between the $s$ and light quark
dynamical masses.

For the $D_s(1^{+'})$ and  $B_s(1^{+'})$  mesons  calculated
masses are also close to those of the  $D$ and $B$ mesons.
Therefore for given  chiral dynamics the $J^P(0^+,1^{+'})$ states
cannot be considered as the chiral partners of the ground-state
multiplet $J^P(0^-,1^-)$, as suggested in \cite{11I}.

We also analyse why two other members of the 1P multiplet, with
$J^P=2^+$ and $1^+$, do not acquire the mass shifts due to decay
channel coupling (DCC) and have small widths. Such situation
occurs if the states $1^+$ and $1^{+'}$ appear to be almost pure
$j=\frac32$ and $j=\frac12$ states. Still small mixing angle
between them, $|\phi|<6^{\circ}$, is shown to be compatible with
experimental restriction on the width of $D_{s1}(2536)$, admitting
possible admixture of other component in the wave function (w.f.)
$\lesssim 10\%$.

In our analysis  the 4-component (Dirac) structure of the light
quark w.f. is crucially important. Specifically, the emission of a
NG meson is accompanied with the $\gamma_5$ factor which permutes
higher and lower components of the Dirac bispinors. For the
$j=1/2,P$ -wave and the $j=1/2,S$ -wave states it is exactly the
case that this ``permuted overlap'' of the w.f. is maximal because
the lower component of the first state is similar to the higher
component of the second state and vice versa. We do not know other
examples of such a ``fine tuning''. On the other hand in the first
approximation we neglect an interaction between two mesons in the
continuum, like $DK$,etc.

In present paper we concentrate on the $P$-wave $B,B_s$ mesons and
the effects of the  channel coupling. While the 1P levels of  the
$D, D_s$ mesons are now established with  good accuracy
\cite{1},\cite{2},\cite{26I}, for the  $B, B_s$  mesons only
relatively narrow $2^+,1^+$ states  have been recently observed
\cite{27I},\cite{28I}. According to these data the splitting
between the $2^+$ and $1^+$ levels is small, $\sim 20-10$ MeV,
while the mass difference between $B_s(2^+)$ and $B(2^+)$ states
is large $\sim 100$ MeV, as for the $D_s(2^+)$ and $D(2^+)$
mesons.

The actual position of the $B(1P),B_s(1P)$ levels is  important
 for several reasons. Firstly, since dynamics  of $(q\bar b)$
mesons is very similar to that of $q\bar c$, the observation of
predicted large mass shifts of  the $B_s(0^+,1^{+'})$ levels would
give a strong argument in favour of  the decay channel mechanism
suggested here and in \cite{18I}. It has been shown in \cite{29a}
that the mass of $B_s(0^+)$ can change by 150 MeV in different
chiral models. Secondly, experimental observation of all $P$-wave
states for the $B$,$B_s$ mesons could  clarify many unclear
features  of spin-orbit and tensor interactions in mesons.
Understanding of  the decay channel coupling (DCC) mass shifts
could become an important step in constructing  chiral theory of
strong decays with emission of one or several NG particles.

\section{Mixing of the $1^+$ and $1^{+'}$ states}

It is well known that  in single-channel approximation, due to
spin-orbit and tensor interactions the $P$-wave multiplet of a HL
meson is split into four levels with $J^P =0^+, 1^+_L, 1^+_H, 2^+$
\cite{29I}. Here  we use the notation H(L) for the higher (lower)
$1^+$ eigenstate of the mixing matrix because a priori one cannot
say which of them mostly consists of the light quark $j=1/2$
contribution. For  a HL meson, strongly coupled to a nearby decay
channel (DC), some member(s) of the $P$-wave multiplet can be
shifted down while another not. Just such situation takes place
for the $D_s(1P)$ multiplet. The position of the levels with
$j=\frac32$, which remains unshifted, will be important in our
analysis.

The scheme of classification, adapted to a HL meson, in the first
approximation treats the heavy quark as a static one and therefore
the Dirac equation can be used to define the light quark levels
and wave functions \cite{10I}. Starting with the Dirac's $P$-wave
levels, one has the states with $j=1/2$ and $j=3/2$. Since the
light quark momentum $j$ and the quantum number $\varkappa$ are
conserved\footnote{we use here the standard notation $ \varkappa
=\mp |j+\frac12|$ for $ j=\left\{\begin{array}{l} l+
\frac12\\l-\frac12\end{array}\right.$}, they run along the
following possible values:

\begin{equation}
\begin{array}{ccc}
\begin{tabular}{c|c|c|c} \multicolumn{4}{c}{even $l$}
\\ \hline $J^P$  & $j$ & $l$ & $\varkappa$ \\ \hline \hline $0^-$
 & ${1}/{2}$ & 0 & -1 \\ \hline $1^-$
 & ${1}/{2}$ & 0 & -1 \\ \hline \hline $1^-$
 & ${3}/{2}$ & 2 & +2
\end{tabular} & \hphantom{aaaaaa} &
\begin{tabular}{c|c|c|c} \multicolumn{4}{c}{ odd $l$}
\\ \hline $J^P$  & $j$ & $l$ & $\varkappa$ \\
\hline \hline $0^+$  & ${1}/{2}$ & 1
& +1 \\ \hline $1^+$  & ${1}/{2}$ & 1 & +1 \\
\hline $1^+$  &
${3}/{2}$ & 1 & -2 \\ \hline $2^+$  & ${3}/{2}$ & 1 & -2 \\
\hline \hline $2^+$  & ${5}/{2}$ & 3 & +3
\end{tabular}
\end{array}
\label{misha_table_3}
\end{equation}

The HL meson w.f. can be expressed in terms of the light quark
w.f. -- the Dirac bispinors $\psi^{jlM}_{q,s}$:
\begin{equation}
\Psi_D\left(J_{1/2}^-,M_f\right)=C^{J,M_f}_{\frac{1}{2},M_f-\frac{1}{2};\frac{1}{2},+\frac{1}{2}}
\psi_q^{\frac{1}{2},0,M_f-\frac{1}{2}}\otimes\bigl|\bar
c\uparrow\bigr\rangle
+C^{J,M_f}_{\frac{1}{2},M_f+\frac{1}{2};\frac{1}{2},-\frac{1}{2}}
\psi_q^{\frac{1}{2},0,M_f+\frac{1}{2}}\otimes\bigl|\bar
c\downarrow\bigr\rangle,
\end{equation}
\begin{equation}
\Psi_{D_s}\left(J_j^+,M_i\right)=C^{J,M_i}_{j,M_i-\frac{1}{2};\frac{1}{2},+\frac{1}{2}}
\psi_s^{j,1,M_i-\frac{1}{2}}\otimes\bigl|\bar
c\uparrow\bigr\rangle
+C^{J,M_i}_{j,M_i+\frac{1}{2};\frac{1}{2},-\frac{1}{2}}
\psi_s^{j,1,M_i+\frac{1}{2}}\otimes\bigl|\bar
c\downarrow\bigr\rangle,
\end{equation}
where
$C^{JM}_{j_1M_1;j_2M_2}$ are the corresponding Clebsch--Gordan
coefficients.

Later in the w.f. we neglect  possible (very small) mixing between
the $D(1^-_{1/2})$, $D(1^-_{3/2})$ states and also between
$D_s(2^+_{3/2})$, $D_s(2^+_{5/2})$ states; however, physical
$D_s(1^+)$ states can be mixed via open channels and tensor
interaction.  The eigenstates, defining the higher $1^+_H$ and
lower $1^+_L$ levels, can be parameterized by introducing the
mixing angle $\phi$:
\begin{equation}
|1^+_H\ran = \cos \phi |j=\frac12\ran +
\sin \phi|j=\frac32\ran\label{5},
\end{equation}
and
\begin{equation}
|1^+_L\ran = -\sin\phi |j=\frac12\ran + \cos\phi
|j=\frac32\ran.\label{6}
\end{equation}
Later we will show that just the $1^+_L$ level with   small
$|\phi|\lesssim 6^{\circ}$, being  almost pure $j=\frac32$  state,
has no DC (hadronic) mass shift. In opposite case  when $1^+_H$ is
mostly $j=\frac32$ state it is convenient  to redefine  in the
equations (\ref{5}), (\ref{6}) the mixing angle as $\phi\to
90^{\circ}-\phi$, performing similar analysis.

In general, the structure of the mixing is   important because it
defines the order of levels,  the  mass shift down of  the
$1^{+'}$ state, as well as the mass shift and the width of another
$1^+$ level. One of our  goals here is to understand why   if the
coupling to nearby continuum channel is taken into account,  the
position of the $2^+$ and $1^+$ levels does not change (within 1-3
MeV)  while $0^+, 1^{+'}$ levels acquire large DC shifts.

\section{Chiral Transitions}

 To obtain the mass shift due to DCC effect we use here
the chiral Lagrangian (\ref{1.1}), which includes both effects of
confinement (embodied in the string tension) and Chiral Symmetry
Breaking (CSB)  (in Euclidean notations): \begin{equation}L_{FCM}
= i\int d^4 x \psi^+ (\hat\partial +m+ \hat M)
\psi\label{14}\end{equation}with the mass operator $\hat M$ given
as a product of the scalar function $W(r)$ and the  SU(3) flavor
octet,

\begin{equation}\hat M = W(r) \exp (i\gamma_5
\frac{\varphi_a\lambda_a}{f_\pi}),\label{15}\end{equation}where
\begin{equation}\varphi_a\lambda_a =\sqrt2
\left(\begin{array}{lll}
\frac{\pi^0}{\sqrt{2}}+\frac{\eta^0}{\sqrt{6}},& \pi^+,& K^+\\
\pi^-,& \frac{\eta^0}{\sqrt{6}}-\frac{\pi^0}{\sqrt{2}},& K^0\\
K^-, &\bar K^0,&
-\frac{2\eta^0}{\sqrt{6}}\end{array}\right).\label{16}\end{equation}Taking
the meson emission to the lowest order, one obtains the quark-pion
Lagrangian in the form \begin{equation}\Delta L_{FCM} =- \int
\psi_i^+ (x) \sigma |\mathbf{x}|\gamma_5
\frac{\varphi_a\lambda_a}{f\pi}
\psi_k(x)d^4x.\label{17}\end{equation}

Writing the equation (\ref{17}) as $\Delta L_{FCM}=- \int V_{if}
dt$, one obtains the operator matrix element for the transition
from the light quark state $i$ (i.e. the initial  state $i$ of a
HL meson) to the continuum state $f$ with the emission of a NG
meson $(\varphi_a\lambda_a)$. Thus we are now able to write the
coupled channel equations, connecting any state of a HL meson  to
a decay channel  which contains another HL meson plus a NG meson.

In  the case, when interaction in each channel and also  in the
transition operator is time-independent, one can write following
system of equations (see \cite{33} for a review)
\begin{equation}[(H_i -E) \delta_{il} +V_{il}]
G_{lf}=1.\label{18}\end{equation}Such two-channel system of the
equations can be reduced to one equation with additional DCC
potential, or the Feshbach potential
\cite{34},\begin{equation}(H_1-E) G_{11} -V_{12} \frac{1}{H_2-E}
V_{21} G_{11}=1.\label{19}\end{equation}Considering a complete set
of  the states $|f\ran$ in the  decay channel 2 and the set of
unperturbed states $|i\ran$ in channel 1, one arrives at the
nonlinear equation for the shifted mass $E$,
\begin{equation}E=E_1^{(i)} - \sum_f \lan i|V_{12} | f\ran
\frac{1}{E_2^{(f)}-E} \lan f
|V_{21}|i\ran.\label{20}\end{equation} Here  the unperturbed
values of $E^{(i)}_1$ are assumed to be known beforehand, while
the interaction $U_{if}$ is defined in (\ref{17}). A solution of
the nonlinear equation (\ref{20}) yields (in general a complex
number $E=\bar E-\frac{i\Gamma}{2})$ one or more roots on all
Riemann sheets of the complex mass plane.

\section{Calculation of the DCC shifts}

To calculate explicitly the  mass shifts, we will use  the Eq.
(\ref{20}) in the following form:
\begin{equation}
m[i]=m^{(0)}[i]-\sum\limits_f \dfrac{|<i|\Hat V|f>|^2}{E_f-m[i]},
\label{21}
\end{equation}
where $m^{(0)}[i]$ is the initial mass, $m[i]$ -- is the final
one, $E_f= \omega_D+\omega_K$ is the energy of the final state,
and the operator $\hat V$ provides the transitions between the
channels (see the comment after Eq. (\ref{17})).

In our approximation we do not take into account the final state
interaction in the $DK$ system and neglect the $D$-meson motion,
so the w.f. of the $i,f$-states are:
\begin{equation}
|f>=\Psi_K(\pp)\otimes\Psi_{D}(M_f),\quad |i>=\Psi_{D_s}(M_i),
\end{equation}
where
\begin{equation}
\Psi_K(\pp)=\frac{e^{i\pp\rr}}{\sqrt{2\omega_K(\pp)}}
\end{equation}
is the plane wave describing the $K$-meson and $\Psi_{D}(M_f)$,
$\Psi_{D_s}(M_i)$ are the HL meson w.f. at rest with the spin
projections $M_f$, $M_i$, respectively.

We introduce the following notations:
\begin{equation}
\omega_K(\pp)=\sqrt{\pp^2+m_K^2},\quad
\omega_D(\pp)=\sqrt{\pp^2+m_D^2},
\end{equation}
so that in the final state the total energy is
$E_f=\omega_D+\omega_K$, while
\begin{equation}
T_f=E_f-m_D-m_K
\end{equation}
is the  kinetic energy. Also it is convenient to define other
masses with respect to nearby threshold: $m_{thr}= m_K+m_D$,
\begin{equation}
E_0=m^{(0)}[D_s]-m_D-m_K,\quad \delta m=m[D_s]-m^{(0)}[D_s],\quad
\Delta = E_0+\delta m=m[D_s]-m_D-m_K,
\end{equation}
where $\Delta$ determines the deviation of the $D_s$ meson mass
from the threshold. In what follows we consider unperturbed masses
$m_0(J^P)$ of the ($Q\bar q$) levels as given (our results do not
change if we slightly vary their position, in this way the
analysis is actually model-independent).

Using these notations, the Eq.(\ref{20}) can be rewritten as
\begin{equation}
E_0-\Delta=\xxi(\Delta), \label{misha_eq_10}
\end{equation}
where
\begin{equation}
\xxi(\Delta)\overset{\text{def}}{=}
\int\frac{d^3\pp}{(2\pi)^3}\sum\limits_{M_f}
\frac{\left|\left\langle M_i\left|\hat
V\right|\pp,M_f\right\rangle\right|^2}{T_f(\pp)-\Delta}
\end{equation} and
\begin{equation}
\left\langle M_i\left|\hat V\right|\pp,M_f\right\rangle=-\int
\Psi^\dag_{D_s}(M_i)\,\sigma
|\rr|\gamma_5\frac{\sqrt{2}}{f_K}\,\Psi_{D}(M_f)\,
\frac{e^{i\pp\rr}}{\sqrt{2\omega_K(\pp)}}\, d^3\rr,
\label{misha_eq_30}
\end{equation}

\begin{figure}
 \caption{Eq.(\ref{misha_eq_10}) for
$E_0<E_0^{\text{crit}}$ (left side) and $E_0>E_0^{\text{crit}}$
(right side)} \label{misha_fig_1-2}
\includegraphics[width=75mm,keepaspectratio=true]{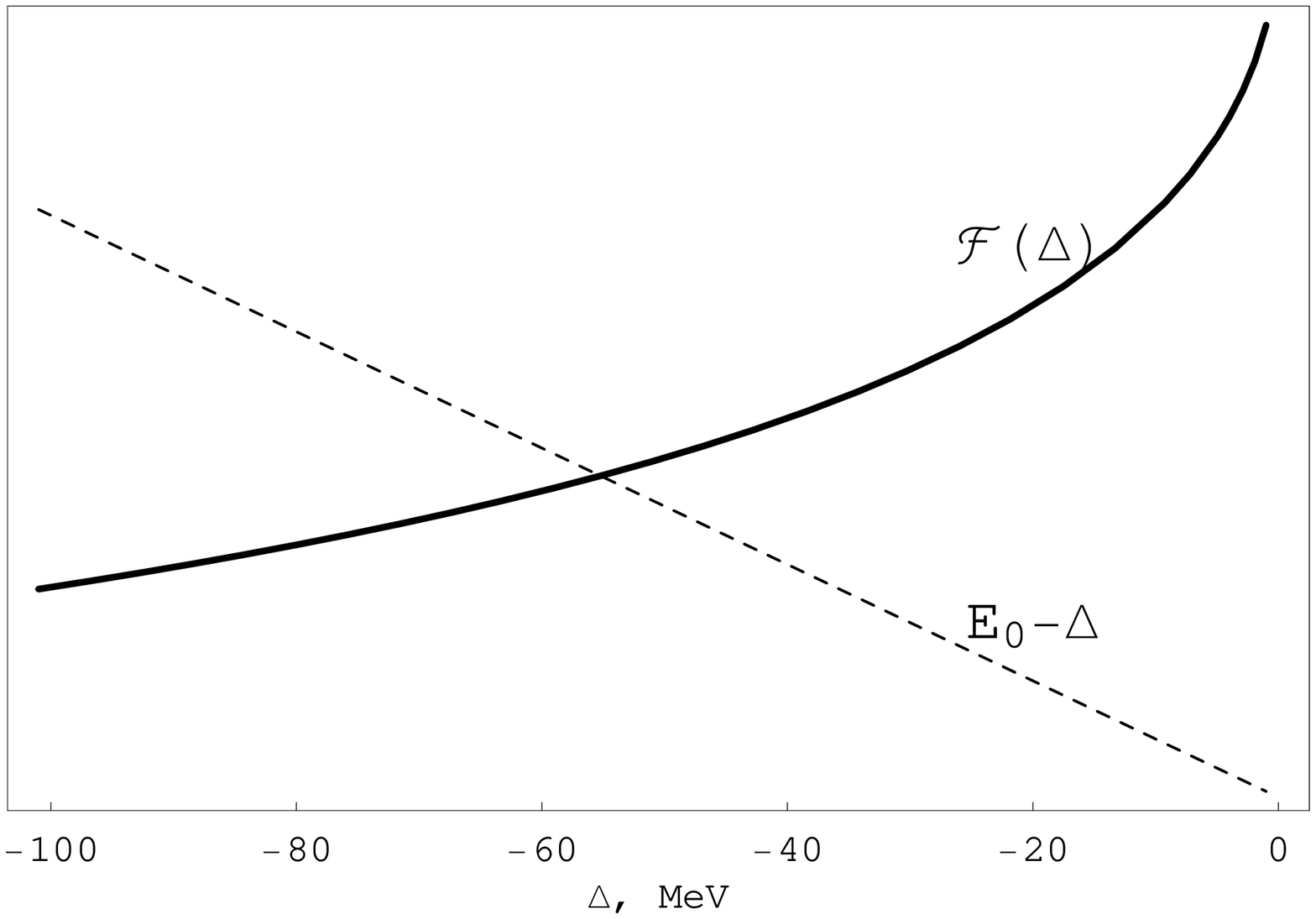}\hspace*{10mm}
\includegraphics[width=75mm,keepaspectratio=true]{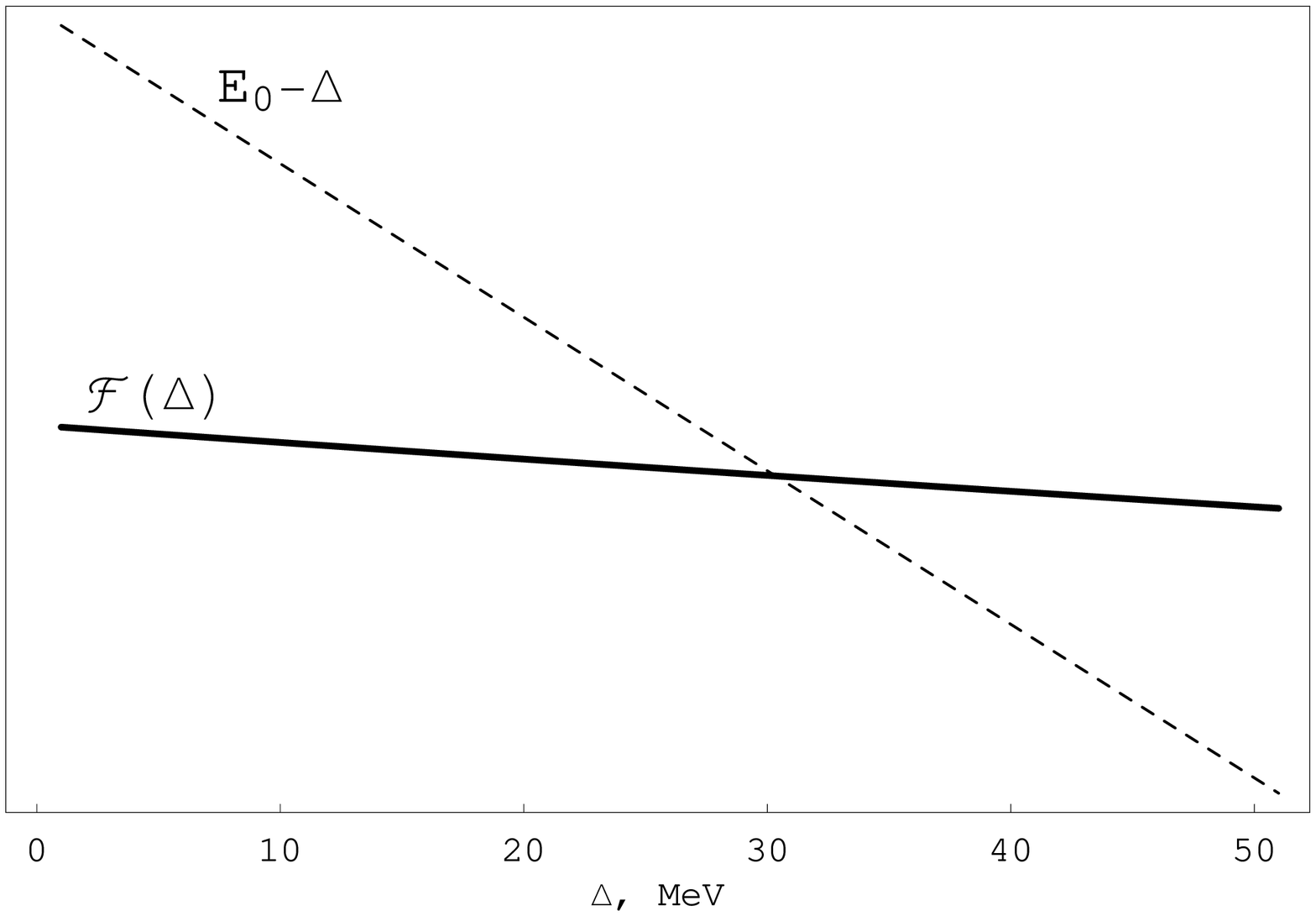}
\end{figure}

The function $\xxi(\Delta)$ for negative $\Delta$ diminishes
monotonously so there exists a final (critical) point,
\begin{equation}
E_0^{\text{crit}}=\xxi(-0).
\end{equation}
Thus, while solving the Eq.(\ref{misha_eq_10}), one has two
possible situations: $E_0<E_0^{\text{crit}}$ and
$E_0>E_0^{\text{crit}}$ (see Fig. \ref{misha_fig_1-2}).

In the first case Eq.(\ref{misha_eq_10}) has a negative real root
$\Delta<0$ and the resulting mass of the $D_s$ meson appears to be
under the threshold. In the second case Eq.(\ref{misha_eq_10}) has
a complex root $\Delta=\Delta'+i\Delta''$ with positive real part
$\Delta'>0$ and negative imaginary part $\Delta''<0$. To find
latter solutions one should make analytic continuation of the
solution(s) from the upper halfplane of $\Delta$ under the cut,
which starts at the threshold, to the lower halfplane (second
sheet). This solution can be also obtained by deforming the
integration contour in $T_f(p)$. In actual calculations we take
infinitesimal imaginary part $\Delta''$, proving that $\Delta$
does not change much for finite $\Delta''$ (the similar procedure
has been used in \cite{18I}). Finally, the resulting mass of the
$D_s$ meson proves to be in the complex plane at the position
$\Delta'-i|\Delta''|$, i.e. the meson has the finite width
$\Gamma=2\Delta''$.

For further calculations  we should insert the explicit meson w.f.
to the matrix element (\ref{misha_eq_30}). As discussed above, in
a HL meson we consider a light quark $q$ moving in the static
field of a heavy antiquark $\bar Q$, and therefore its w.f. can be
taken  as the Dirac bispinor:
\begin{equation} \psi_q^{jlM}=\begin{pmatrix}
g(r)\Omega_{jlM}\\
(-1)^{\frac{1+l-l'}{2}}f(r)\Omega_{jl'M}\end{pmatrix},\quad
\int\limits_0^\infty \left(f^2+g^2\right)r^2dr=1,\end{equation}
where the functions $g(r)$ and $f(r)$ are the  solutions of the
Dirac equation: \begin{equation}
\begin{gathered} g'+\frac{1+\varkappa}{r}g-\left(E_q+m_q+U-V_C\right)f=0,\\
f'+\frac{1-\varkappa}{r}f+\left(E_q-m_q-U-V_C\right)g=0.
\end{gathered} \end{equation} Here the interaction between the quark
and the antiquark is described by a sum of  linear scalar
potential and the  vector Coulomb potential with
$\alpha_s=\text{const}$:
\begin{equation}
U=\sigma r,\quad V_C=-\dfrac{\beta}{r},\quad
\beta=\dfrac{4}{3}\alpha_s.
\end{equation}

Introducing new dimensionless variables
\begin{equation}
x=r\sqrt{\sigma},\quad \varepsilon_q=E_q/\sqrt{\sigma},\quad
\mu_q=m_q/\sqrt{\sigma}, \end{equation} and new dimensionless
functions
\begin{equation} g=\sigma^{3/4}\frac{G(x)}{x},\quad
f=\sigma^{3/4}\frac{F(x)}{x},\quad \int\limits_0^\infty
\left(F^2+G^2\right)dx=1, \end{equation} we come to the following
system of equations:
\begin{equation}
\begin{gathered}
G'+\frac{\varkappa}{x}G-\left(\varepsilon_q+\mu_q+x+\frac{\beta}{x}\right)F=0,
\\
F'-\frac{\varkappa}{x}F+\left(\varepsilon_q-\mu_q-x+\frac{\beta}{x}\right)G=0.
\end{gathered}
\end{equation}This system has been solved numerically.

\begin{figure}
\caption{$G_{1,2,3}(x)$ functions (left side) and  $F_{1,2,3}(x)$
functions (right side) } \label{misha_fig_3-4}
\includegraphics[width=75mm,keepaspectratio=true]{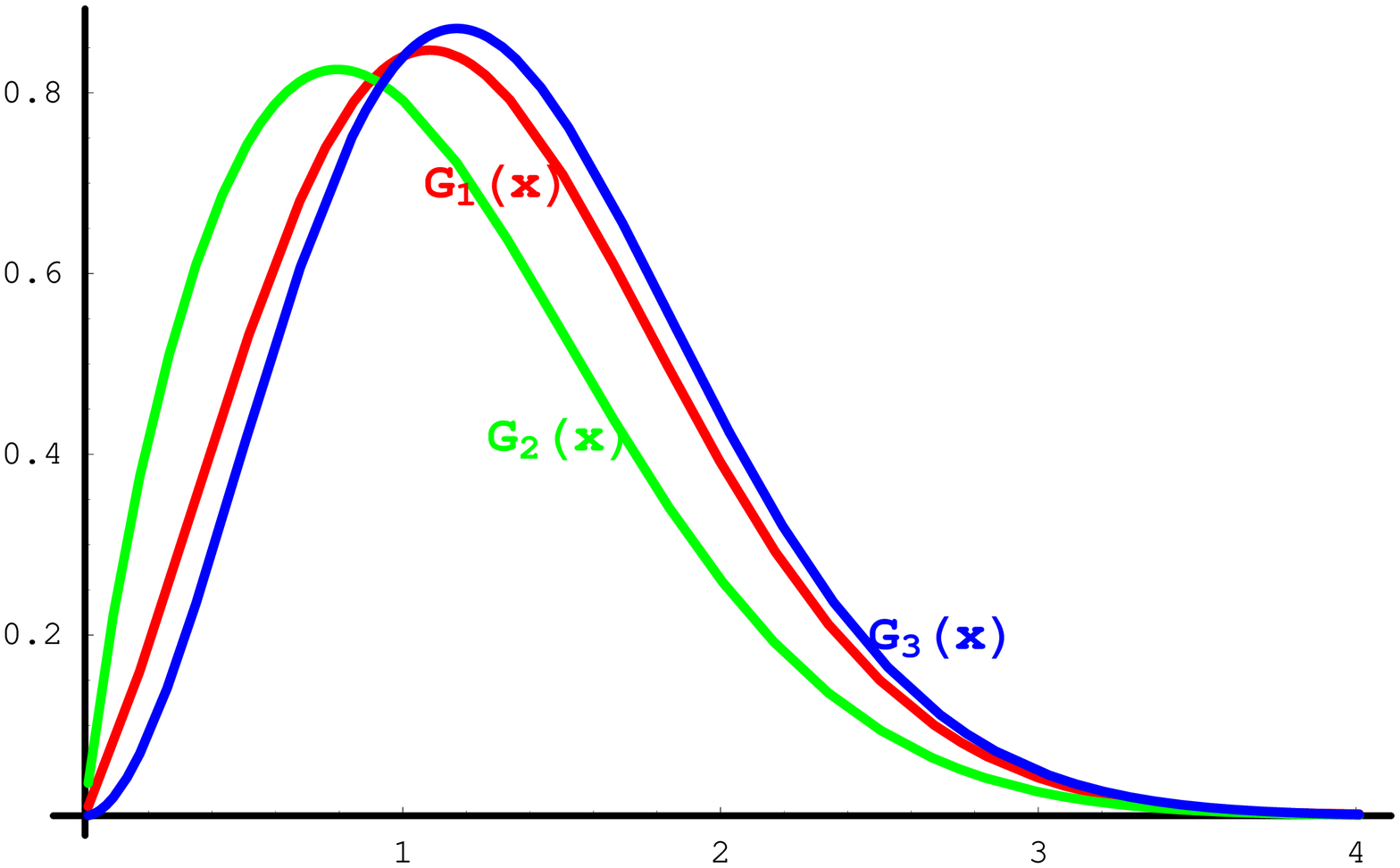}\hspace*{10mm}
\includegraphics[width=75mm,keepaspectratio=true]{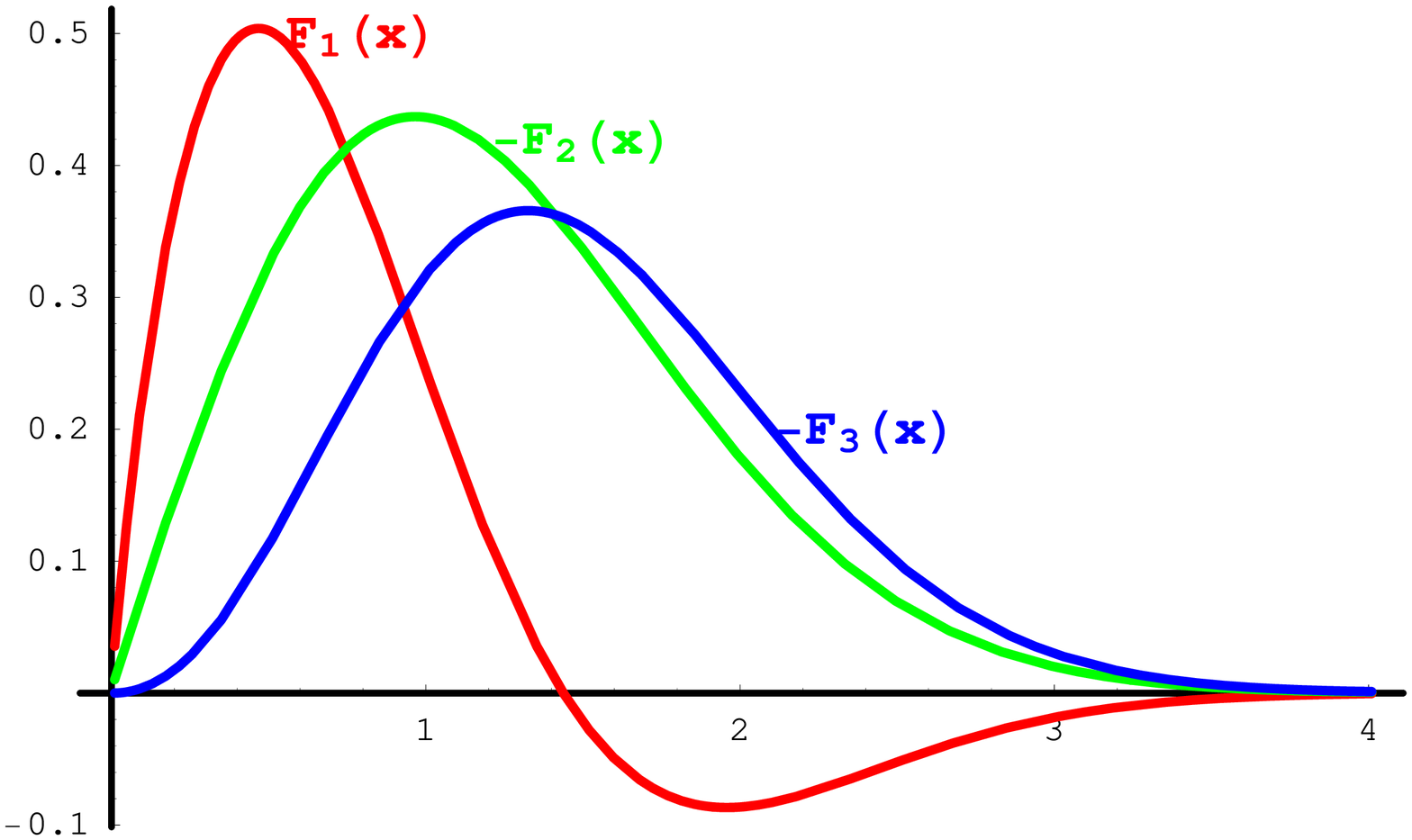}
\end{figure}

Using the parameters from the papers \cite{dirac}:
\begin{equation} \begin{gathered} \sigma=0.18~\text{GeV}^2,\quad \alpha_s=0.39,
\\ m_s=210~\text{MeV},\quad m_q=4~\text{MeV}, \end{gathered}
\end{equation}
we obtain the following Dirac eigenvalues $\varepsilon$:
\begin{equation}
\begin{tabular}{r|l|l} $\varkappa$ &
$\bar Qq$, $\mu_q=0.01$~ & $\bar Qs$, $\mu_s=0.5~$ \\
\hline\hline -1 & 1.0026 & 1.28944 \\ \hline +1 & 1.7829 & 2.08607 \\
\hline -2 & 1.7545 & 2.08475
\end{tabular}
\label{misha_table_2}
\end{equation}
and corresponding eigenfunctions $G$, $F$ are given in Fig.
\ref{misha_fig_3-4}.

Our choice of $\sigma$ and $\alpha_s$ is a common one in the frame
of the FCM approach, and the value of the light quark mass really
does not influence here on any physical results because of its
smallness in comparison with the natural mass scale
$\sqrt{\sigma}$. The strange quark mass is taken from \cite{ms},
where it was found from the ratio of experimentally measured decay
constants $f(D_s)/f(D)$; the same value can be obtained by a
renormalization group evolution starting from the conventional
value $m_s(\text{2~GeV}) = 90\pm 15\text{~GeV}$.

Later we use the simplified notations for the quark bispinors:
\begin{equation}
\psi_1(M_1)\overset{\text{def}}{=}\psi_s^{\frac{1}{2},1,M_1},\quad
\psi_2(M_2)\overset{\text{def}}{=}\psi_q^{\frac{1}{2},0,M_2},\quad
\psi_3(M_3)\overset{\text{def}}{=}\psi_s^{\frac{3}{2},1,M_3}.
\end{equation}

\begin{figure}
\caption{$\Phi_{0,2}(q)$ functions} \label{misha_fig_5}
\centerline{\includegraphics[width=100mm,keepaspectratio=true]{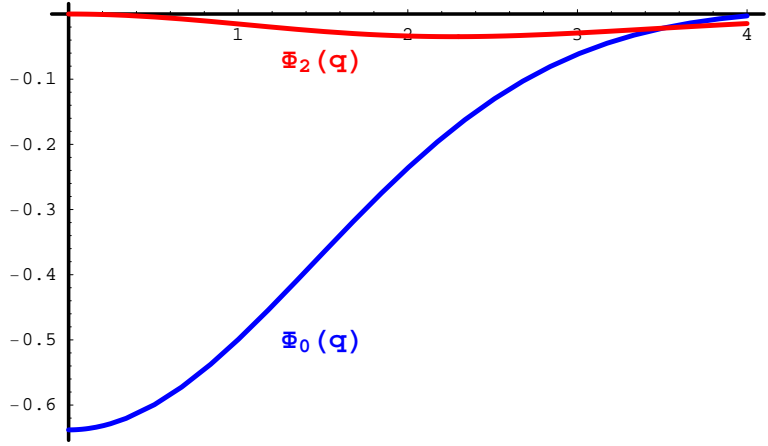}}
\end{figure}

Now, using  explicit expressions for the spherical spinors,
\begin{equation}
\Omega_{l+1/2,l,M}=\begin{bmatrix}
\sqrt{\frac{j+M}{2j}}\,Y_{l,M-1/2} \vphantom{\bigg|} \\
\sqrt{\frac{j-M}{2j}}\,Y_{l,M+1/2} \vphantom{\bigg|}
\end{bmatrix},\qquad \Omega_{l-1/2,l,M}=\begin{bmatrix}
-\sqrt{\frac{j-M+1}{2j+2}}\,Y_{l,M-1/2} \vphantom{\bigg|} \\
\sqrt{\frac{j+M+1}{2j+2}}\,Y_{l,M+1/2} \vphantom{\bigg|}
\end{bmatrix}, \end{equation}
and the expansion :
\begin{equation} e^{i\pp\rr}=4\pi\sum\limits_{l,M} i^l j_l(pr)
Y^*_{l,M}\left(\frac{\pp}{p}\right)
Y_{l,M}\left(\frac{\rr}{r}\right)\, , \end{equation} after
cumbersome transformations (which are omitted in the text) we
obtain the transition matrix elements:
\begin{equation} \Bigl\|\mathcal{V}_{12}\Bigr\|_{M_1,M_2}=-\int
\psi^\dag_1(M_1)\,\sigma
|\rr|\gamma_5\frac{\sqrt{2}}{f_K}\,\psi_2(M_2)\,
\frac{e^{i\pp\rr}}{\sqrt{2\omega_K(\pp)}}\, d^3\rr
=\frac{\sqrt{\sigma}}{f_K\sqrt{\omega_K(p)}}\Phi_0\left(\frac{p}{\sqrt{\sigma}}\right)
\sqrt{4\pi}Y^*_{0,M_1-M_2}\left(\frac{\pp}{p}\right),
\end{equation}
\begin{multline} \Bigl\|\mathcal{V}_{32}\Bigr\|_{M_3,M_2}=-\int
\psi^\dag_3(M_3)\,\sigma
|\rr|\gamma_5\frac{\sqrt{2}}{f_K}\,\psi_2(M_2)\,
\frac{e^{i\pp\rr}}{\sqrt{2\omega_K(\pp)}}\, d^3\rr\\
=-\frac{\sqrt{\sigma}}{f_K\sqrt{\omega_K(p)}}\Phi_2\left(\frac{p}{\sqrt{\sigma}}\right)
\sqrt{\frac{4\pi}{5}}Y^*_{2,M_3-M_2}\left(\frac{\pp}{p}\right)
\times \begin{bmatrix} -1 & +2 \\ -\sqrt{2} & +\sqrt{3} \\
-\sqrt{3} & +\sqrt{2} \\ -2 & +1
\end{bmatrix}\, .
\end{multline}
where
\begin{equation} \begin{gathered} \Phi_0(q)=\int\limits_0^\infty j_0(qx)xdx\Bigl[
G_1(x)F_2(x)-F_1(x)G_2(x) \Bigr], \\
\Phi_2(q)=\int\limits_0^\infty j_2(qx)xdx\Bigl[
G_3(x)F_2(x)-F_3(x)G_2(x) \Bigr]. \end{gathered} \end{equation}
Notice that because of different signs of the $F_1(x)$ and
$F_{2,3}(x)$ functions (while the $G_{1,2,3}$ functions are all
positive) on almost all real axis, the integral $\Phi_2$ appears
to be strongly suppressed in comparison with the integral
$\Phi_0$. This fact is confirmed by numerical simulations (see
Fig. \ref{misha_fig_5}).

Finally, introducing universal functions
\begin{equation}
\begin{gathered}
\tilde\xxi_{0,2}(\Delta)=\frac{\sigma}{2\pi^2f_K^2}\int\limits_0^\infty
\frac{p(T_f)\omega_D(T_f)dT_f}{T_f+m_D+m_K}\cdot
\frac{\Phi_{0,2}^2\left(\dfrac{p(T_f)}{\sqrt{\sigma}}\right)}{T_f-\Delta}\,
,
\\ \tilde\Gamma_{0,2}(T_f)=\frac{\sigma}{\pi f_K^2}\cdot
\frac{p(T_f)\omega_D(T_f)}{T_f+m_D+m_K} \cdot
\Phi_{0,2}^2\left(\frac{p(T_f)}{\sqrt{\sigma}}\right)\, ,
\end{gathered}
\end{equation}
we come to the following equations to determine meson masses and
widths:

\begin{equation}
\begin{array}{rc||cl} D_s(0^+) \vphantom{\bigg|} &
\hphantom{a} & \hphantom{a} &
E_0[0^+]-\Delta=\tilde\xxi_0(\Delta),
\\ D_s(1^+_L) \vphantom{\bigg|} & & & E_0[1^+_L]-\Delta=
\cos^2\phi\cdot\tilde\xxi_0(\Delta)+\sin^2\phi\cdot\tilde\xxi_2(\Delta),
\\ D_s(1^+_H)
\vphantom{\bigg|} & & & E_0[1^-_H]-\Delta'=
\sin^2\phi\cdot\tilde\xxi_0(\Delta')+\cos^2\phi\cdot\tilde\xxi_2(\Delta'),
\\ \vphantom{\bigg|} & & &
\Gamma[1^+_H]=\sin^2\phi\cdot\tilde\Gamma_0(\Delta')
+\cos^2\phi\cdot\tilde\Gamma_2(\Delta'),
\\ D_s(2^+_{3/2})
\vphantom{\bigg|} & & & E_0[2^+_{3/2}]-\Delta'=
\dfrac{3}{5}\cdot\tilde\xxi_2(\Delta'),
\\ \vphantom{\bigg|} & & &
\Gamma[2^+_{3/2}]=\dfrac{3}{5}\cdot\tilde\Gamma_2(\Delta').
\end{array}
\label{misha_table_4}
\end{equation}

\section{Results and discussion}

In this chapter, using the expressions (\ref{misha_table_4}) to
define the $D_s$ and $B_s$ meson mass shifts, we present and
discuss our results. We will take into account the following pairs
of mesons in coupled channels ($i$ refers to first (initial)
channel, while $f$ refers to second (decay) one):

\begin{equation}
\begin{array}{c}
\begin{tabular}{c||c|c|c} $i$ & $D_s(0^+)$ &
$D_s(1^+)$ & $D_s(2^+)$
\\ \hline $f$ & $D(0^-)+K(0^-)$ & $D^*(1^-)+K(0^-)$ &
$D^*(1^-)+K(0^-)$
\end{tabular}\\[7mm]
\begin{tabular}{c||c|c|c} $i$ & $B_s(0^+)$ &
$B_s(1^+)$ & $B_s(2^+)$
\\ \hline $f$ & $B(0^-)+K(0^-)$ & $B^*(1^-)+K(0^-)$ &
$B^*(1^-)+K(0^-)$
\end{tabular}
\end{array}
\label{misha_table_1}
\end{equation}

In our calculations we use  the following meson masses and
thresholds (in MeV):
\begin{equation}
\begin{gathered}
m_{D^+}=1869, \quad m_{D^+}+m_{K^-}=2363, \\
m_{D^{*+}}=2010, \quad m_{D^{*+}}+m_{K^-}=2504,\\
m_{B^+}=5279, \quad m_{B^+}+m_{K^-}=5772, \\
m_{B^*}=5325, \quad m_{B^*}+m_{K^-}=5819. \\
\end{gathered}
\end{equation}

\begin{table}
\caption{$D_s(0^+)$-meson mass shift due to the $DK$ decay channel
and $B_s(0^+)$-meson mass shift due to the $BK$ decay channel (all
in MeV)} \label{misha_table_11}
\begin{center}
\begin{tabular}{|c|c|c|c|c|} \hline state  & $m^{(0)}$ &
$m^{\text{(theor)}}$ & $m^{\text{(exp)}}$ & $\delta m$  \\
\hline \hline $D_s(0^+)$  & 2475 (30)& 2330(20)& 2317 & -145
\\ \hline $B_s(0^+)$ & 5814(15) & 5709 (15) & {\footnotesize not seen} & -105 \\
\hline
\end{tabular}
\end{center}
\end{table}

\begin{table}
\caption{The $D_s(1^+)$, $D_s(2^+)$ meson mass shifts and widths
due to the $D^*K$ decay channel for the mixing angle $4^\circ$
(all in MeV)} \label{misha_table_12}
\begin{center}
\begin{tabular}{|c|c|c|c|c|c|c|} \hline state & $m^{(0)}$ &
$m^{\text{(theor)}}$ & $m^{\text{(exp)}}$ &
$\Gamma^{\text{(theor)}}_{(D^*K)}$ & $\Gamma^{\text{(exp)}}_{(D^*K)}$ & $\delta m$ \\
\hline \hline $D_s(1^+_H)$ & 2568(15) & 2458(15) & 2460 & $\times$
& $\times$ & -110
\\ \hline $D_s(1^+_L)$ & 2537 & 2535(15) & 2535(1) & 1.1 & $<1.3$ &
-2  \\
\hline $D_s(2^+_{3/2})$ & 2575 & 2573 & 2573(2) & 0.03 &
{\footnotesize
not seen} & -2\\
\hline
\end{tabular}
\end{center}
\end{table}

\begin{table}
\caption{The $B_s(1^+)$, $B_s(2^+)$ meson mass shifts and widths
due to the $B^*K$ decay channel for the mixing angle $4^\circ$
(all in MeV)} \label{misha_table_13}
\begin{center}
\begin{tabular}{|c|c|c|c|c|c|c|} \hline state & $m^{(0)}$ &
$m^{\text{(theor)}}$ & $m^{\text{(exp)}}$ &
$\Gamma^{\text{(theor)}}_{(B^*K)}$ & $\Gamma^{\text{(exp)}}_{(B^*K)}$ & $\delta m$ \\
\hline \hline $B_s(1^+_H)$ & 5835(15) & 5727 & {\footnotesize not
seen} & $\times$ & $\times$ & -108
\\  \hline $B_s(1^+_L)$ & 5830(fit) & 5828 & 5829 (1)& 0.8 & $<2.3$ &
-2
\\
\hline $B_s(2^+_{3/2})$ & 5840(fit) & 5838 & 5839(1) & $<10^{-3}$
&
{\footnotesize not seen} & -2 \\
\hline
\end{tabular}
\end{center}
\end{table}


The results of our calculations are presented in Tables
\ref{misha_table_11}--\ref{misha_table_13}. \textit{A priori} one
cannot say whether the $|j=\frac12\ran$ and $|j=\frac32\ran$
states are mixed or not. If there is no mixing at all, in this
case the width $\Gamma(D_{s1}(2536))= 0.3$ MeV is obtained in
\cite{35}, while the experimental limit is $\Gamma<2.3$ MeV
\cite{26I} and recently in \cite{36} the width $\Gamma=1.0\pm
0.17$ MeV has been measured. Therefore small mixing is not
excluded and here we take the mixing angle $\phi$ slightly
deviated from  $\phi=0^{\circ}$ ( no mixing case). Then we define
those angles $\phi$ which are compatible with experimental data
for the  masses and widths of both $1^+$ states.

\begin{figure}
\caption{Schemes of $D_s(1^+,2^+)$ and $B_s(1^+,2^+)$ shifts due
to chiral coupling} \label{misha_fig_6-7}
\includegraphics[width=75mm,keepaspectratio=true]{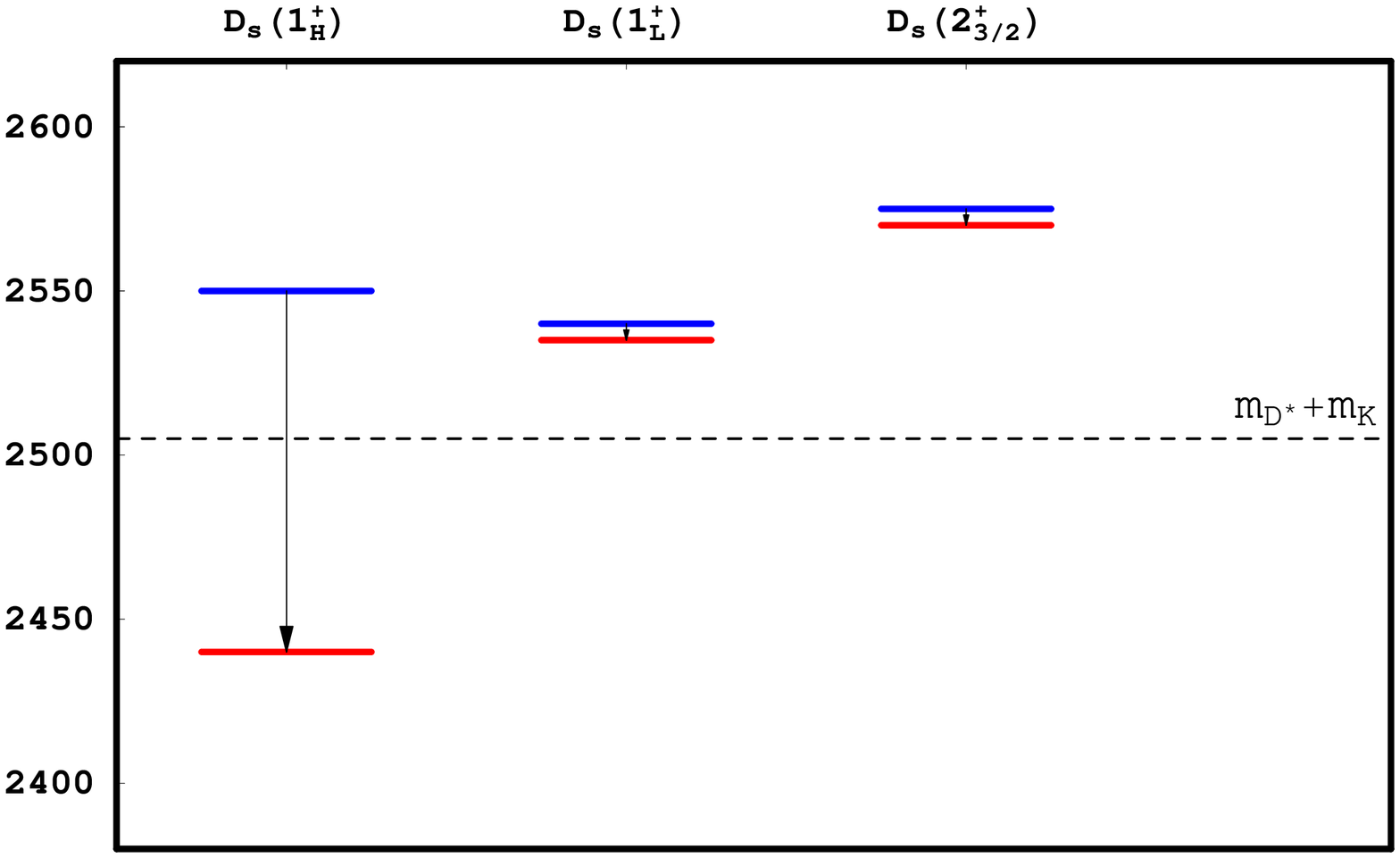}\hspace*{10mm}
\includegraphics[width=75mm,keepaspectratio=true]{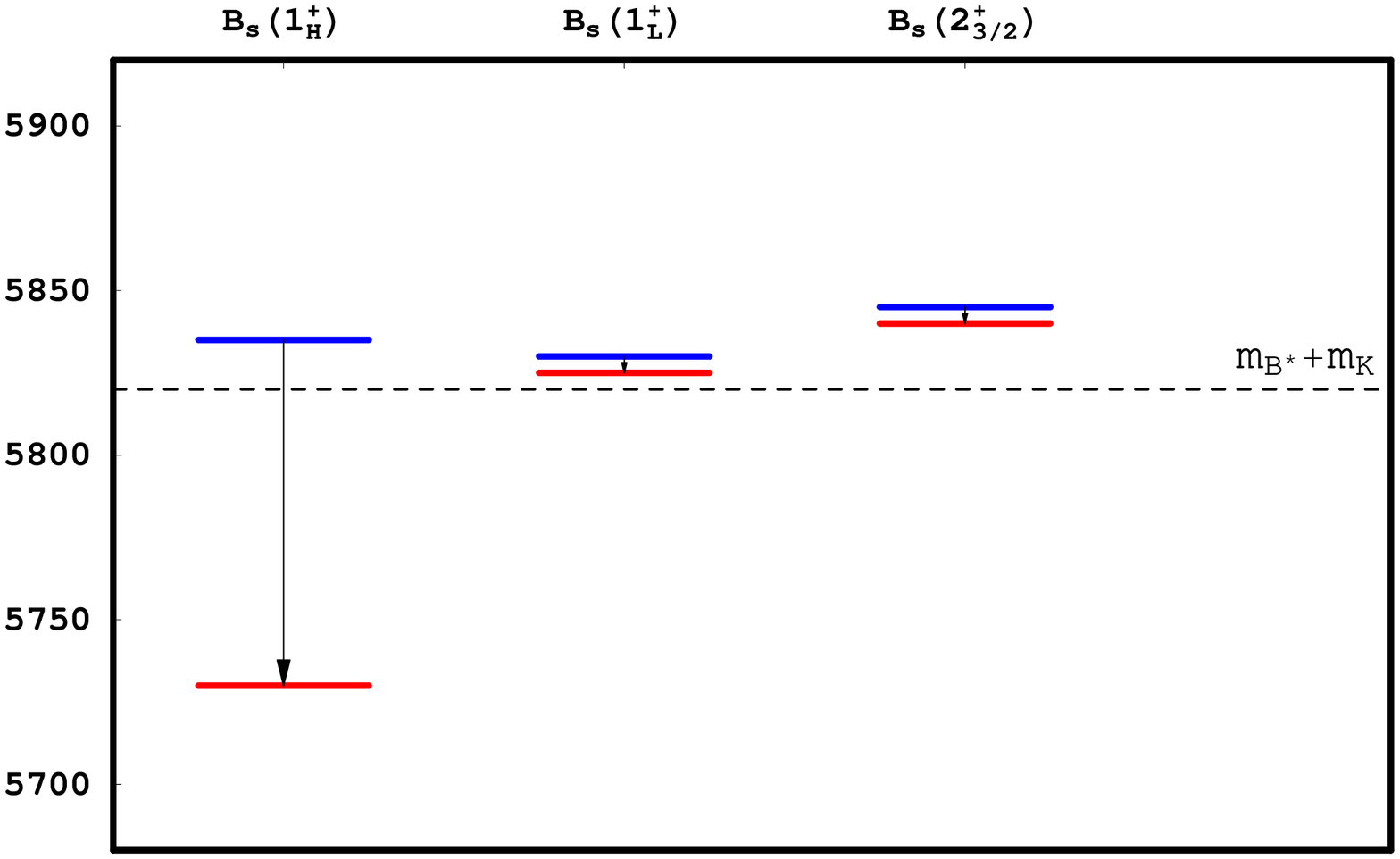}
\end{figure}

The small value $\phi= 5.7^{\circ}$ provides large mass shift
($\sim 100$ MeV) of the for the $1_{H}^+(j=1/2)$ level and at the
same time does not produce the mass shift of the $1^{+}_L$ level,
which is almost pure $j=\frac32$ state. For illustration we show
the scheme of the $1^+$, $2^+$ shifts on Fig. \ref{misha_fig_6-7}.
We would like to stress here that the mass shifts weakly differ
for $D_s$ and $B_s$,
 or weakly depend on the   heavy quark mass: this  can be directly
illustrated using in the Eq.(\ref{misha_table_4}) the expansion
via the inverse heavy quark mass.

\section{Conclusions}

We have studied the mass shifts of the $D_s(0^+, 1^{+'})$ and
$B_s(0^+, 1^{+'})$ mesons due to strong coupling to the decay
channels $DK, D^*K$ and $BK, B^*K$. To this end the chiral
quark-pion Lagrangian without fitting parameters has been used.

We have shown that the emission of a NG meson, accompanied with
the $\gamma_5$ factor, gives rise to maximal overlapping  between
the higher component with $j=\frac12$ of the $P$-wave meson
($D_s,B_s$) bispinor w.f. and the lower component (also with
$j=\frac12$) of the $S$-wave HL meson w.f. in considered $S$-wave
decay channel. Due to this effect, while taking the w.f. of the
$1P$ and $1S$ states with the use of the Dirac equation, large
mass shifts of the $0^+, 1^{+'}$ states are obtained. In
particular, the shifted masses $M(B_s,0^+)=5710(15)$ MeV and
$M(B_s,1^{+'})=5730(15)$ MeV were calculated in agreement with the
predictions in \cite{14I} and of S.Narison \cite{9I} and by $\sim
100$ MeV lower than in \cite{3I},\cite{4I},\cite{10I}.

The widths of $D_{s1}(2536)$ and $B_{s1}(5830)$ are also
calculated. To satisfy the experimental condition
$\Gamma(D_{s1}(2536))<2.3$ MeV the following limit on the mixing
angle $\phi$ (between the $|j=\frac32>$ and $|j=\frac12>$ states)
is obtained: $|\phi|\lesssim 6^{\circ}$.

\section*{Acknowledgments}
The authors would like to acknowledge support from the President
Grant No. 4961.2008.2 for scientific schools. One of the authors
(M.A.T.) acknowledges partial support from the President Grant No.
MK-2130.2008.2 and the RFBR for partial support via Grant No.
06-02-17120.

\end{document}